\documentclass[aps,prb,twocolumn,showpacs,preprintnumbers,amsmath,amssymb,superscriptaddress]{revtex4}

\usepackage{graphicx}
\usepackage{dcolumn}
\usepackage{bm}

\begin{document}

\title{Spin-polarized Andreev transport influenced by Coulomb repulsion through two quantum dot system}

\author{Piotr Trocha}
\email{piotrtroch@gmail.com}\affiliation{Department of Physics,
Adam Mickiewicz University, 61-614 Pozna\'n, Poland}

\author{J\'ozef Barna\'s}
 \affiliation{Department of Physics, Adam
Mickiewicz University, 61-614 Pozna\'n, Poland}
\affiliation{Institute of Molecular Physics, Polish Academy of
Sciences, 60-179 Pozna\'n, Poland}

\date{\today}

\begin{abstract}

Spin-polarized transport through a double quantum dot system attached to a common superconducting lead and two ferromagnetic electrodes (fork geometry) is investigated theoretically. The key objective of the analysis is to describe the influence of electrodes' ferromagnetism on the Andreev tunneling. Both direct and crossed Andreev tunneling  processes are considered, in general. The other objective is a detailed analysis of the role of Coulomb interaction and its impact on the Andreev tunneling processes.

\end{abstract}

\maketitle

\section{Introduction}{\label{Sec:1}}

Due to the presence of a superconducting gap $\Delta$ in the
density of states of a superconductor, single electron tunneling processes to the superconductor are
blocked for bias voltages smaller than the energy gap. However,
the electrons can tunnel from the normal metal (or ferromagnetic metal) into
the superconductor {\it via} Andreev-like
processes.~\cite{andreev,tinkham}.
Originally Andreev reflection  was discovered in junctions of normal metals
and superconductors. An electron with energy
$\epsilon<\Delta$ incident on the interface of the junction from
the normal metal can not propagate into the superconductor as there
are no available quasiparticle states within the energy gap. Thus,
the electron must be reflected at the interface. Such a process is
known as the normal reflection. However, the reflection can also occur in
a different way. At the interface, the incident electron can form a Cooper pair with another electron of opposite spin and wavevector, which propagates into the superconductor, while a hole is reflected  back
to the normal metal.
This reflection is known as the Andreev reflection (AR).

Analogous processes can occur in a quantum dot coupled to
normal and superconducting leads (N-QS-S).~\cite{claughton,yeyati,deacon} Here, an
electron with energy $\epsilon$ tunnels from the normal metal lead
through discrete dot's levels. However, it can not tunnel alone
into the superconductor when its energy is smaller than the energy gap,
$|\epsilon|<\Delta$. Such an electron, however,  can pick up another
electron with energy $-\epsilon$ and opposite spin to create a
Cooper pair which is able to propagate into the superconductor.
During this process, a hole with energy $-\epsilon$ and spin opposite
to that of the incident electron is reflected back into
the normal-metal lead.

There exists a vast literature on the Andreev tunneling in
various quantum dot systems attached to one normal and one
superconducting
leads.~\cite{fazioPRL98,sunQDSC,WangPRB,donabidowicz07,domanski08,panSC2,
zhangSC,pengPRB05,tanakaSC,governale08}
However, Andreev transport through ferromagnet-QD-superconductor
(F-QD-S) structures is rather in an early stage of investigations~\cite{feng03,cao,pengZhang,wysokinskiJPCM}.
From theoretical side, mostly equilibrium
regime was considered and the on-site Coulomb correlations were neglected.
From experimental side, in turn, fabricating such a system in reality and
performing proper measurements is a real challenge, and to our knowledge, there exist up to now only one experimental work
dealing with a quantum dot coupled to ferromagnetic and
superconducting leads.~\cite{csonka}

Attaching an additional normal-metal electrode to the N-QD-S system can lead to a new
kind of Andreev tunneling processes. These processes are
analogous to the crossed Andreev reflections (CAR) described theoretically~\cite{deutscher02,deutscherAPL00}
and observed experimentally in hybrid structures consisting of two
normal-metal  (or ferromagnetic) leads connected {\it via} point contacts to a common superconducting
electrode.~\cite{beckmannPRL04,russoPRL05} In
contrast to the direct AR, where the hole is reflected back to the
electrode from which the incoming electron arrives, in the CAR
processes the hole is reflected into the second, spatially separated
electrode. However,  the distance
between the two contacts should be smaller than the size of Cooper
pairs formed in the superconducting lead (more precisely, smaller than the corresponding coherence length).

A single quantum dot coupled to two normal and one
superconducting leads gives the possibility of investigating not only the crossed
Andreev reflection,~\cite{sunSC01,chenAPL04,feng06,golubev07,koning09} but also
a Cooper pair splitting (which can be understood as
the effect inverse to  CAR). However, more efficient Cooper pair splitting
can be achieved in a system based on two
quantum dots.~\cite{hofstetterSC,herrmann10SC} The advantage of double quantum dot (DQD) systems follows from
the possibility of independent tuning of the dots' levels. Such a dot's level tuning is important as
finite-bias experiments have shown that Cooper pair
splitting can be dominant in resonance, whereas out
of the resonance elastic cotunelling processes
dominate.~\cite{hofstetterPRL11,kleineEPL09,bursetPRB11}
It is also worth to note that creation of nonlocal entangled electrons
has been earlier proposed and investigated theoretically in a similar
DQD and multi-dot systems.~\cite{lossPRB01,martinPRB04,martinPRB06}
This entanglement can be probed by measuring noise cross
correlations.~\cite{melinPRB08,chevallierPRB11}

\begin{figure}
\begin{center}
\includegraphics[width=0.28\textwidth,angle=-90]{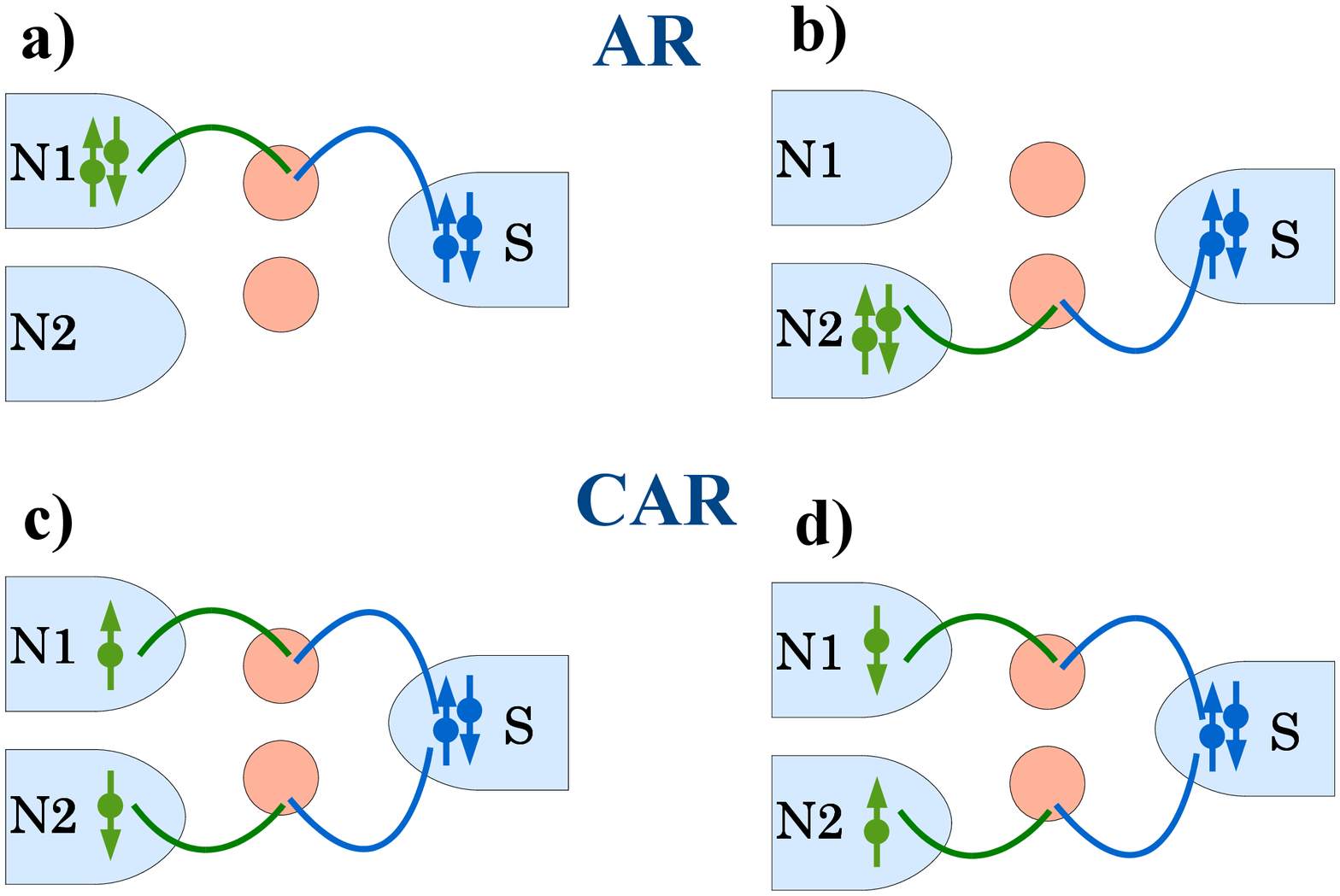}\caption{\label{Fig:1} (color on-line) Schematic
picture of direct (a-b) and crossed (c-d) Andreev tunneling  in a
double quantum dot system in the fork geometry. N1 and N2 are two normal-metal leads (ferromagnetic in general) and S is a superconducting lead.}
\end{center}
\end{figure}

From the above follows that systems based on  DQDs have some advantages over those based on single dots.
Therefore, in this paper we  study transport in the system based on a DQD connected to ferromagnetic and superconducting leads.
Generally, the influence of ferromagnetism on Andreev transport
through DQD system is only weakly explored.~\cite{liJP08,cabreraPRB10}
This also applies to the influence of
Coulomb correlations in DQD structures on Andreev
tunneling~\cite{cabreraPRB10}, which is rather unexplored even in the case of nonmagnetic
leads.~\cite{tanakaSC,lambert,konigPRB11,baranski}
We consider the system  in the fork geometry, as shown schematically in Fig.~\ref{Fig:1}, and analyze
the Andreev tunneling processes. We distinguish between processes
 in which only a single dot is involved ({\it
direct Andreev tunneling} (DAT)~[Fig.~\ref{Fig:1} (a-b)]), and
Andreev tunneling events in which both dots participate (called
{\it crossed Andreev tunneling} (CAT) [Fig.~\ref{Fig:1} (c-d)]). There are also virtual process which do not lead to creation (or annihilation) of Cooper pairs in the superconductor. The process, called {\it elastic cotunneling}, transfer electron between two normal-metal leads {\it via} virtual states in the superconductor. We are especially interested in two aspects of Andreev tunneling. The first one is the question how spin polarization of electrons taking part in transport affects the Andreev tunneling. The second aspect is the role of Coulomb interaction on the dots.

In section 2 we present the theoretical background. This includes description of the  model and also description of the method used to
calculate the current. Numerical results are presented and discussed in section 3.
Final conclusions are in section 4.

\section{Theoretical background}

\subsection{Model}

We consider electron transport through the system consisting of
two quantum dots (QD1 and QD2) coupled to a common superconducting electrode.
Additionally, each of the dots is attached to a separate ferromagnetic lead,
as shown schematically in Fig.~\ref{Fig:1}.
The ferromagnetic leads are assumed to be in a collinear magnetic
configuration: either parallel or antiparallel. The DQD system can be
then modeled by the following Hamiltonian:
\begin{equation}\label{Eq:1}
H_{DQD}=\sum_{i\sigma}\varepsilon_{i\sigma}d^\dag_{i\sigma}d_{i\sigma}+\sum_i
   U_in_{i\sigma}n_{i\bar{\sigma}},
   \end{equation}
where $\varepsilon_{i\sigma}$ and $U_i$  denote the spin dependent dots' energy level and
Coulomb integral for the dot $i$ ($i=1,2$).  The leads'
Hamiltonian has the form: $H_{\rm leads}=H_L +H_R$, with $H_L$
describing ferromagnetic leads in the non-interacting
quasi-particle approximation, and $H_R$ denoting the BCS
Hamiltonian of the superconducting lead. In the mean field approximation
$H_R$ takes the form:
\begin{eqnarray}\label{Eq:2}
H_{R}&=&\sum_{{\mathbf
k}}\limits\sum_\sigma\limits\varepsilon_{{\mathbf k}R\sigma}
     c^\dag_{{\mathbf k}R\sigma}c_{{\mathbf k}R\sigma}
     \nonumber\\
     &+&
     \sum_{{\mathbf k}}\limits\left(\Delta^{\ast} c_{{\mathbf
k}R\downarrow}c_{{-\mathbf k}R\uparrow}+\Delta c^{\dag}_{{-\mathbf
k}R\uparrow}c^{\dag}_{{\mathbf k}R\downarrow}\right),
\end{eqnarray}
with $\varepsilon_{{\mathbf k}R\sigma}$ denoting the relevant
single-particle energy and  $\Delta=|\Delta|e^{i\Phi}$ standing
for the  order parameter of the superconductor. Here, $|\Delta|$
denotes the superconducting gap, whereas $\Phi$ is the relevant
phase. In the problem considered here there is only one superconducting lead, so
the phase factor is irrelevant. Thus, we will omit this factor in
further consideration, and , and
assume $\Delta$ is real and positive. 

Finally, spin conserving electron tunneling between the leads and the dots
is described by the Hamiltonian,
\begin{equation}\label{Eq:2b}
   H_{\rm T}=\sum_{\mathbf{k}j}\limits\sum_{i\sigma}
   \limits V_{i\mathbf{k}\sigma}^{Lj} c^\dag_{\mathbf{k}j\sigma}d_{i\sigma}
   +\sum_{\mathbf{k}}\limits\sum_{i\sigma}
   \limits V_{i\mathbf{k}\sigma}^{R} c^\dag_{\mathbf{k}R\sigma}d_{i\sigma}
     + \rm
   H.c..
\end{equation}
The first term describes coupling of the dots to the (left) ferromagnetic leads, with $V_{i\mathbf{k}\sigma}^{Lj}$ being the matrix elements
of tunneling between the $i$th ($i=1,2$) dot and the $j$th ($j=1,2$) ferromagnetic lead. In turn, the second term presents coupling to the supperconducting (right) lead with $V_{i\mathbf{k}\sigma}^R$ denoting the relevant tunneling matrix elements.

Coupling of the $i$th  dot to the $j$th ferromagnetic lead will be parameterized in terms of
$\Gamma_{Lji}^{\sigma}=2\pi\langle |V^{Lj}_{i}|^2\rangle\rho_{Lj}^\sigma$, where all the spin dependence is captured by the spin-dependent density of states $\rho_{Lj}^\sigma$ in the $j$th lead, and $\langle |V^{Lj}_{i}|^2\rangle$ is the independent of spin average value of tunneling matrix elements.
We write these parameters in the form
$\Gamma_{L11}^{\sigma}=\beta\Gamma_L(1+\tilde{\sigma}p)$,
$\Gamma_{L22}^{\sigma}=\Gamma_L(1\pm\tilde{\sigma}p)$,
with $\tilde{\sigma}=1$ for $\sigma=\uparrow$, and
$\tilde{\sigma}=-1$ for $\sigma=\downarrow$, and the upper (lower)
sign in $\Gamma_{L22}^{\sigma}$ corresponding to the parallel
(antiparallel) magnetic configuration. We assume that magnetic moment of the left lead corresponding to $j=2$ is reversed in the antiparallel configuration.
Here, $p$ is the spin polarization of the density of states at the Fermi level in the ferromagnetic leads, and $\beta$ describes asymmetry in the coupling of the two dots to the two ferromagnetic leads.
We also assume there is no coupling between the $j$th lead and $i$th dot for $j\ne i$,
$\Gamma_{L12}^{\sigma}=\Gamma_{L21}^{\sigma}=0$.
Similarly, one can parameterize the coupling of the superconducting
lead to the dots by the parameters $\Gamma_{Rij}^{\sigma}=2\pi\langle |V^R_{i}V^R_{j}|\rangle\rho_{R}$, where $\rho_{R}$ is the independent of spin density of states (per spin) in the normal state of the right lead.
We write  the coupling parameters as $\Gamma_{R11}^{\sigma}=\alpha\Gamma_R$,
$\Gamma_{R22}^{\sigma}=\Gamma_R$,
$\Gamma_{R12}^{\sigma}=q_s\Gamma_R\sqrt{\alpha}$. In a general
case, $\alpha$ takes into account difference in the coupling of the superconducting  electrode to the two dots, and $q_s$ describes strength of the nondiagonal elements. Moreover, the DQD system can be
asymmetrically coupled to the left and right leads.
To take this into account we introduce the asymmetry parameter
$r$ defined as $\Gamma_R/ 2\Gamma_L=r$. Apart from this, we define $\Gamma$ as   $\Gamma=2\Gamma_L.$

\subsection{Current}

The current flowing through the system can be obtained in terms of the nonequilibrium Green's
function formalism.
In the expanded basis (Nambu representation~\cite{nambu}) this formula has the form~\cite{zheng}:
\begin{eqnarray}\label{Eq:3}
  J &=&\frac{ie}{2\hbar}\sum_\sigma\limits{\rm  Tr} \int\frac{d\varepsilon}{2\pi}
  \left\{[\mathbf{\Gamma}_{L}^\sigma-\mathbf{\Gamma}_{R}^\sigma]
  \mathbf{G}^{<}_\sigma(\varepsilon)\right.
  \nonumber\\
  &+&
  \left.[\mathbf{f}_L(\varepsilon)\mathbf{\Gamma}_{L}^\sigma-
   \mathbf{f}_R(\varepsilon)\mathbf{\Gamma}_{R}^\sigma]
   [\mathbf{G}^{r}_\sigma(\varepsilon)-\mathbf{G}^{a}_\sigma(\varepsilon)]\right\}.
\end{eqnarray}
Here, $\mathbf{G}^{r,a,<}_\sigma(\varepsilon)$ are the Fourier
transforms of the retarded, advanced, and lesser Green's functions in the matrix form, while
$\mathbf{\Gamma}_L^{\sigma}$ and $\mathbf{\Gamma}_R^{\sigma}$ denote the coupling matrices
to the left and right leads, respectively. More specifically, the coupling matrix to the left (ferromagnetic) leads takes the form,
\begin{equation}\label{Eq:11}
  \mathbf{\Gamma}_{L}^\sigma=
  \Gamma_L
  \left[%
\begin{array}{cccc}
  \beta (1+\tilde{\sigma}p) & 0 & 0  & 0 \\
  0 & \beta (1-\tilde{\sigma}p) & 0 & 0 \\
  0 & 0 &  1\pm \tilde{\sigma}p & 0 \\
  0 & 0 & 0 & 1\mp \tilde{\sigma}p \\
\end{array}%
\right],
\end{equation}
where the upper and lower signs correspond to the parallel and antiparallel magnetic configurations of the left ferromagnetic leads, respectively.
The coupling matrix to the right (superconducting) lead, in turn, can be written in the form~\cite{zheng}
\begin{equation}\label{Eq:12}
  \frac{\mathbf{\Gamma}_{R}^\sigma}{\rho_R}=
    \Gamma_R
  \left[%
\begin{array}{cccc}
  \alpha & - \tilde{\sigma}\frac{\Delta}{|\varepsilon|}\alpha & q_s\sqrt{\alpha}
  & - \tilde{\sigma}\frac{\Delta}{|\varepsilon|}q_s\sqrt{\alpha} \\
  - \tilde{\sigma}\frac{\Delta}{|\varepsilon|}\alpha & \alpha
  & - \tilde{\sigma}\frac{\Delta}{|\varepsilon|}q_s\sqrt{\alpha} & q_s\sqrt{\alpha} \\
  q_s\sqrt{\alpha} & - \tilde{\sigma}\frac{\Delta}{|\varepsilon|}q_s\sqrt{\alpha} & 1
  & - \frac{\Delta}{|\varepsilon|} \\
  - \tilde{\sigma}\frac{\Delta}{|\varepsilon|}q_s\sqrt{\alpha} & q_s\sqrt{\alpha}
   & - \tilde{\sigma}\frac{\Delta}{|\varepsilon|} & 1 \\
\end{array}%
\right],
\end{equation}
where $\rho_R=\rho_R(\varepsilon )$  is a normalized BCS density of states,
defined as the ratio of the density of states in the superconducting
phase and the density of states in the normal state,
\begin{equation}\label{Eq:13}
\rho_R(\varepsilon)=\frac{|\varepsilon|\theta(|\varepsilon|-\Delta)}
{\sqrt{\varepsilon^2-\Delta^2}}.
\end{equation}
Note, $\rho_R(\varepsilon )$ (and thus also $\mathbf{\Gamma}_{R}^\sigma$) vanishes in the superconducting gap.
Finally,
$\mathbf{f}_{L}(\epsilon)$ in Eq.(4) is the matrix of relevant
Fermi-Dirac distribution functions for the left leads,
\begin{equation}\label{Eq:14}
\mathbf{f}_{L}(\epsilon)=
\left[%
\begin{array}{cccc}
  f(\varepsilon-eV_1) & 0 & 0 & 0 \\
  0 & f(\varepsilon+eV_1) & 0 & 0 \\
  0 & 0 & f(\varepsilon-eV_2) & 0 \\
  0 & 0 & 0 & f(\varepsilon+eV_2) \\
\end{array}%
\right],
\end{equation}
where $V_j$ ($j=1,2$) is the electrostatic potential applied to the $j$-th ferromagnetic electrode, and $f(\varepsilon)$ is the Fermi-Dirac distribution.
In turn, the matrix $\mathbf{f}_{R}(\varepsilon)$ for
the superconducting lead has the form $\mathbf{f}_{R}(\varepsilon)=f(\varepsilon)\,{\rm diag}(1,1,1,1)$.

To derive the lesser Green's function we apply the Keldysh relation,
\begin{eqnarray}
\mathbf{G}^{<}_\sigma (\varepsilon )=\mathbf{G}^{r}_\sigma (\varepsilon )\mathbf{\Sigma}^{<}_\sigma \mathbf{G}^{a}\sigma (\varepsilon ),\label{Eq:4}
\end{eqnarray}
whereas the lesser self-energy can be obtained from the following
formula
\begin{eqnarray}
\mathbf{\Sigma}^{<} =\mathbf{\Sigma}^{<}_{L }+\mathbf{\Sigma}^{<}_{R  }=
i[\mathbf{f}_L(\varepsilon)\mathbf{\Gamma}_{L}^\sigma +
   \mathbf{f}_R(\varepsilon)\mathbf{\Gamma}_{R}^\sigma ],\label{Eq:5}
\end{eqnarray}
which is valid in the absence of interactions in the central
region of the device. This equation also holds for interactions
taken in the mean field approximations. Making use of Eqs.
(\ref{Eq:4}) and (\ref{Eq:5}) and taking into account the
identity,
$\mathbf{G}^{r}_\sigma (\varepsilon)-\mathbf{G}^{a}_\sigma (\varepsilon)=-i\mathbf{G}^{r}_\sigma (\varepsilon)\mathbf{\Gamma}^\sigma \mathbf{G}^{a}_\sigma (\varepsilon)$,
with $\mathbf{\Gamma}^\sigma =\mathbf{\Gamma}_{L}^\sigma +\mathbf{\Gamma}_{R}^\sigma $,
the current expression (\ref{Eq:3}) reduces to the
Landauer-like formula. When the bias voltage is applied to the
system with equal electrochemical potentials of the ferromagnetic
leads, $\mu_{L1}=\mu_{L2}=eV$, and the electrochemical potential of the
superconducting electrode is equal to zero, $\mu_R=0$, the current acquires
the following form:
\begin{equation}\label{Eq:6}
  J=J^S+J^{A},
\end{equation}
with $J^S$ and $J^{A}$ defined as follows:
\begin{eqnarray}
J^S=&&\frac{e}{h}\sum_\sigma\limits\int d\varepsilon [f_L(\varepsilon-eV)-f_R(\varepsilon)]
\nonumber\\
&&\times \sum_{i=1,3}\limits
[\mathbf{G}^{r}_\sigma(\varepsilon)\mathbf{\Gamma}_{R}^\sigma\mathbf{G}^{a}_\sigma(\varepsilon)\mathbf{\Gamma}_{L}^\sigma]_{ii},\label{Eq:7}
\\
J^A=&&\frac{e}{h}\sum_\sigma\limits\int d\varepsilon [f_L(\varepsilon-eV)-f_L(\varepsilon+eV)]
\nonumber\\
&&\times \sum_{i=1,3,}^{j=2,4}\limits
G^r_{ij\sigma}(\varepsilon)[\mathbf{\Gamma}_{L}^\sigma\mathbf{G}^{a}_\sigma(\varepsilon)\mathbf{\Gamma}_{L}^\sigma]_{ji}.\label{Eq:8}
\end{eqnarray}
In the above equation, $J^A$ denotes the current due to {\it Anndreev
tunneling}, whereas $J^S$ includes contributions from  the following processes: (i)
usual tunneling of electrons with energy $|\varepsilon|>\Delta$, (ii)
{\it branch crossing} processes in which an electron from the
normal lead is converted into the hole like state in the superconducting
lead, (iii) processes in which electron (hole) tunnels from left
lead into the superconductor picking up the quasiparticle
(quasihole) and creating (annihilating) a Cooper pair. At  $T=0$ K
and $|eV|<\Delta$, the term $J^S$ vanishes, $J^S=0$, and the only
contribution to  current originates from the {\it Andreev}
reflection. However, both components contribute for $|eV|>\Delta$.
The tunneling processes giving rise to $J^A$ and
$J^S$ (except  single-quasiparticle tunneling) occur through virtual states
of the dots.

The retarded Green's function $\mathbf{G}^{r}_\sigma(\varepsilon)$ has
been obtained from Dyson equation
\begin{equation}\label{Eq:9}
\mathbf{G}^{r}_\sigma(\varepsilon)=[(\mathbf{g}^r_\sigma(\varepsilon))^{-1}+\mathbf{\Sigma}^r_\sigma]^{-1},
\end{equation}
where $\mathbf{g}^r_\sigma(\varepsilon)$ denotes the Fourier transform of the retarded
Green's function of the two dots isolated from the leads, and
$\mathbf{\Sigma}^r_\sigma =\mathbf{\Sigma}^r_{L\sigma}+\mathbf{\Sigma}^r_{R\sigma}$ is the retarded self-energy due to
interaction between the DQD system and electrodes. The retarded Green's
function  $\mathbf{g}^{r}_\sigma$ has been derived from the relevant
equation of motion (see the Appendix A).
The retarded self-energy
$\mathbf{\Sigma}^r_{L\sigma}$,
taken in the wide band approximation,
can be written in a simple form as
$\mathbf{\Sigma}^{r}_{L\sigma}=-\frac{i}{2}\mathbf{\Gamma}_L^\sigma $, with
$\mathbf{\Gamma}_L^\sigma$  denoting the coupling matrix of the DQD system
to the left (ferromagnetic) leads (see Eq.(5)).
On the other hand, the self-energy due to coupling of the dots to the superconducting lead has the following form:
$\mathbf{\Sigma}^{r}_{R\sigma}=  -\frac{i}{2}[ \tilde{\rho}_R(\varepsilon)/\rho_R(\varepsilon)] \mathbf{\Gamma}_R^\sigma$,
where $\mathbf{\Gamma}_R^\sigma$ is given by Eq.(6), while  $\tilde{\rho}_R(\varepsilon)$ denotes a dimensionless  modified
BCS density of states of the superconductor of the following form:
\begin{equation}\label{Eq:13}
\tilde{\rho}_R(\varepsilon)=\frac{|\varepsilon|\theta(|\varepsilon|-\Delta)}
{\sqrt{\varepsilon^2-\Delta^2}} -i
\frac{\varepsilon\theta(\Delta-|\varepsilon|)}
{\sqrt{\Delta^2-\varepsilon^2}}.
\end{equation}
Note, that the real part of  $\tilde{\rho}_R(\varepsilon)$ vanishes inside the superconducting gap, while the imaginary part vanishes outside the gap.
It is also worth to mention that $\mathbf{\Gamma}_{R}^\sigma$ can be expressed as $\mathbf{\Gamma}_{R}^\sigma=i(\mathbf{\Sigma}^{r}_{R\sigma}-\mathbf{\Sigma}^{a}_{R\sigma})$, where $\mathbf{\Sigma}^{a}_{R\sigma}=[\mathbf{\Sigma}^{r}_{R\sigma}]^\dag$.  .

\section{Numerical results}

Numerical calculations have been performed for both linear and
nonlinear response regimes. To concentrate only on the Andreev current, we
assume $T=0$ K and $|eV_i|<\Delta$ (assuming the same electrostatic potentials for
both ferromagnetic leads, $V_1=V_2=V$). All energy quantities have
been expressed in the units of $\Delta$.
Furthermore, we assume $\Gamma=0.1$, $U_1=U_2\equiv U$,  and
$\varepsilon_1=\varepsilon_2\equiv\varepsilon_0$.
\begin{figure}
\begin{center}
\includegraphics[width=0.45\textwidth,angle=0]{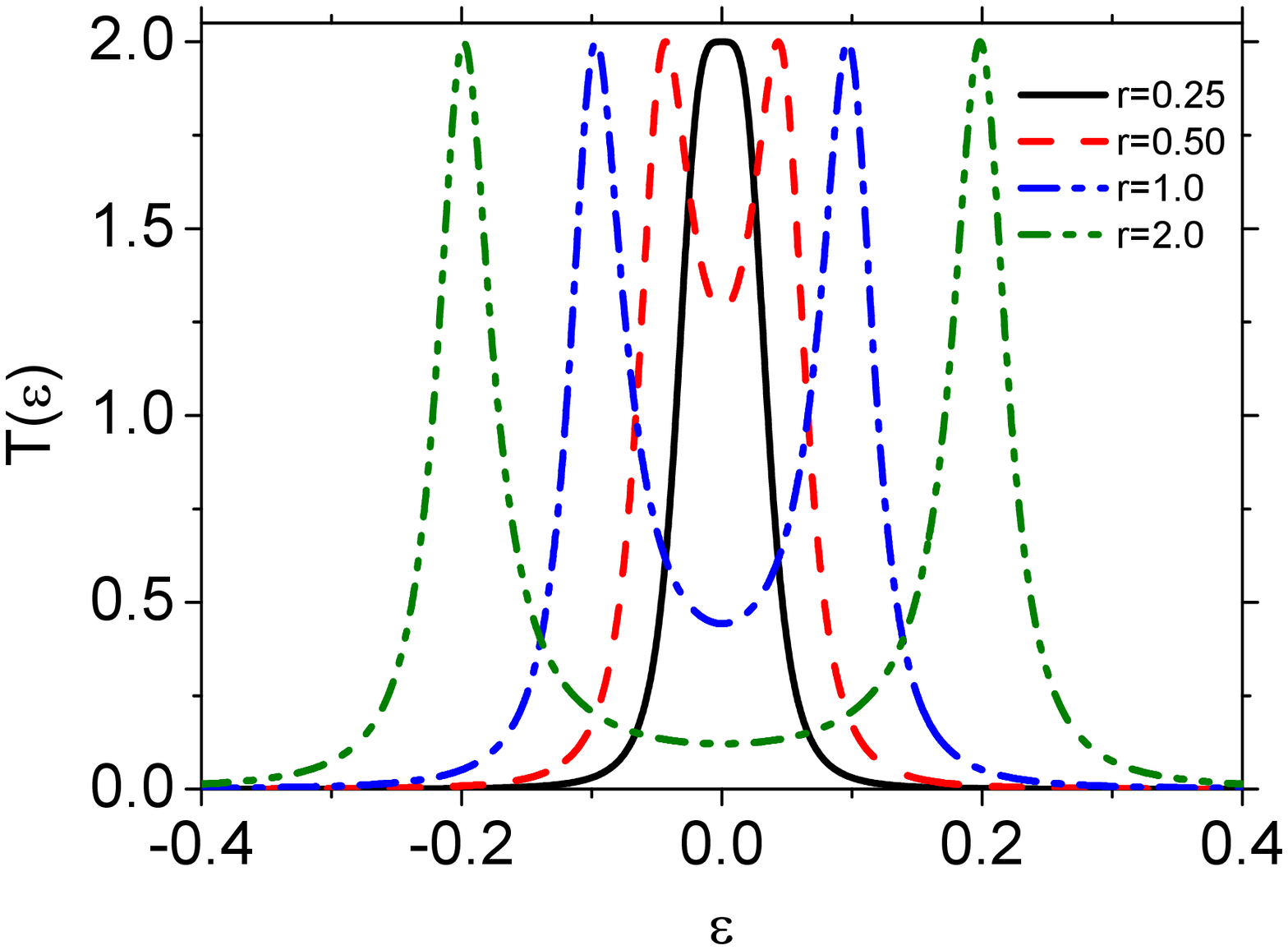}
\caption{\label{Fig:2}
(color on-line) Transmission function at equilibrium, calculated for indicated values of the asymmetry parameter  $r$. The other parameters are: $p=0$, $q_s=1$,
$\alpha=\beta=1$, $\varepsilon_0=0$,  $U=0$.}
\end{center}
\end{figure}

\subsection{Nonmagnetic leads, $p=0$}
To clarify the influence of magnetic polarization of the leads on  Andreev transport through the system, we consider first the case of nonmagnetic left electrodes, $p=0$.

\subsubsection{Vanishing Coulomb repulsion, $U=0$}
For the sake of simplicity, we start the description of numerical results from
the situation when the on-site Coulomb interaction is absent, $U=0$. This interaction will be restored later.
Figure~\ref{Fig:2} shows the equilibrium transmission function $T(\varepsilon)$ through the DQD system
in the fork geometry, calculated for indicated values of the
asymmetry parameter $r$ and for $q_s=1$. The transmission function is given by the formula,
\begin{equation}
T(\varepsilon)=\sum_\sigma\limits
\sum_{i=1,3,}^{j=2,4}\limits
G^r_{ij\sigma}(\varepsilon)[\mathbf{\Gamma}_{L}^\sigma\mathbf{G}^{a}_\sigma(\varepsilon)\mathbf{\Gamma}_{L}^\sigma]_{ji},
\end{equation}
see also Eq.(13).
One can notice that
for a symmetric case, $\Gamma_L=\Gamma_R$ (corresponding to $r=0.5$), the two peaks in the
transmission are not fully resolved, whereas for $r>0.5$ the peaks
are very well separated. This can be explained when realizing that
coupling to the superconducting lead gives rise to a renormalization of the dots' energy
levels. Due to the electron-hole symmetry, the density of states
of each dot reveals two peaks situated at
$\pm\sqrt{\varepsilon_0^2+\Gamma_R^2}$ (corresponding to two
Andreev states). In turn, coupling to the ferromagnetic
electrodes gives rise  to a broadening of these resonances of the order of
$\Gamma_L$. Thus, when increasing strength of the coupling  to the
superconducting electrode,  the two peaks move away from each
other. For $\Gamma_R<\Gamma_L$, the broadening of the two
resonances is relatively large and one observes only one maximum,
see the curve for $r=0.25$ in Fig.~\ref{Fig:2}.
It is worth noting that coupling to the superconductor does not
contribute to the broadening of Andreev states~\cite{claughton}, because the corresponding coupling matrix vanishes,
$\mathbf{\Gamma}_R^\sigma=0$, and only real part of the self-energy
$\mathbf{\Sigma}_{R\sigma}$ is nonzero for
energies within the energy gap.

The results presented above were obtained for the case when the two dots
were coupled to the superconducting lead with equal strengths, $\alpha=1$, and the nondiagonal coupling elements (responsible for indirect coupling between the dots {\it via} the superconducting lead) were maximal, $q_s=1$.
The situation with $q_s=1$ can occur when the distance between contacts of the superconducting lead to the dots is smaller than the corresponding superconducting coherence length. Thus, increase in the distance between the contacts may be modeled by decreasing value of the parameter $q_s$. Finally, the distance between the contacts for $q_s=0$ is much longer than the coherence length, and the CAT processes are fully suppressed.
\begin{figure}
\begin{center}
\includegraphics[width=0.43\textwidth,angle=0]{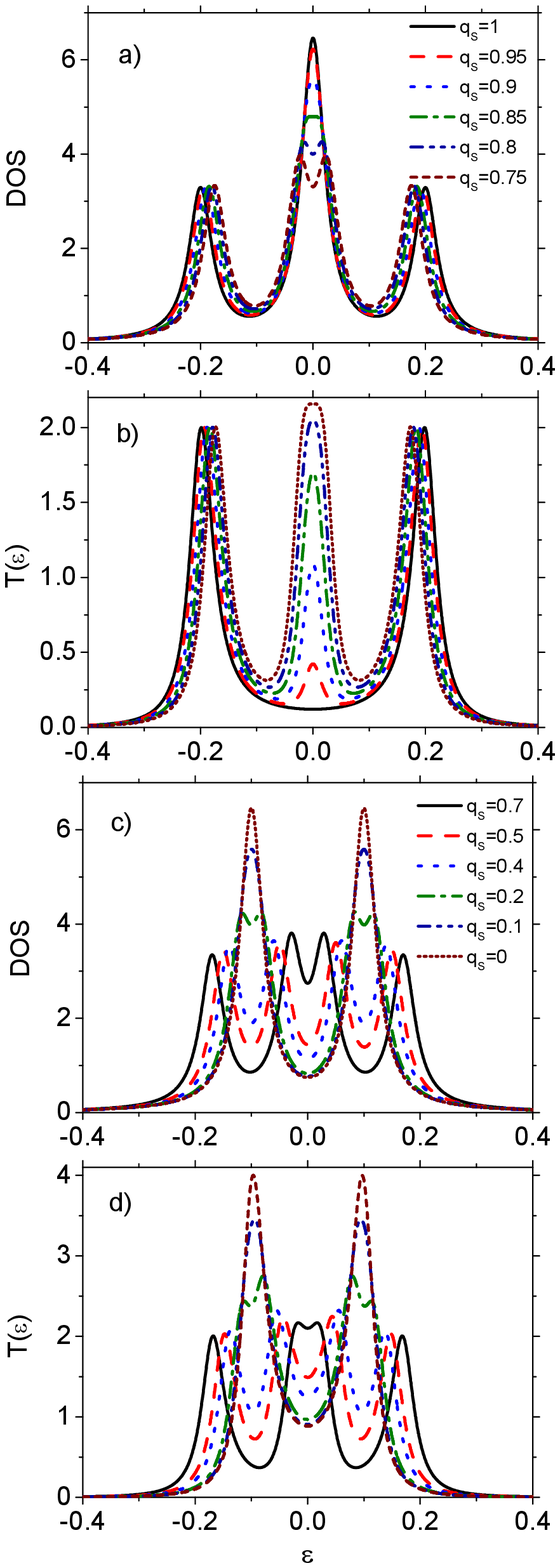}
\caption{\label{Fig:3}
(color on-line) Equilibrium density of states [(a) and (c)] and the corresponding transmission function [(b) and (d)] calculated for indicated values of the parameter  $q_s$. The other parameters are: $p=0$, $r=2$,
$\alpha=\beta=1$, $\varepsilon_0=0$, $U=0$.}
\end{center}
\end{figure}

Figure~\ref{Fig:3} shows the transmission function at equilibrium and the corresponding density of states of the two dots for indicated values of the parameter $q_s$ and $\alpha=1$. The particle density of states was calculated as ${\rm DOS}=-(1/\pi ){\rm Im} \{G^r_{11\sigma}$, and
assume  is real and positive. Generally, the density of states exhibits four peaks attributed to the four Andreev bound states, which emerge in the DQD system coupled to the superconducting lead. For $\varepsilon_0=0$, the maxima in the density of states corresponding to the Andreev bound states are  at the energies, $\varepsilon_{+}=\pm\Gamma_R [(1+\alpha )/2+\sqrt{(\alpha-1)^2/4+q_s^2\alpha}]/2$ and $\varepsilon_{-}=\pm\Gamma_R [(1+\alpha )/2-\sqrt{(\alpha-1)^2/4+q_s^2\alpha}]/2$. For $q_s=0$ (and $\alpha=1$), only two peaks in the density of states appear, which are located at $\varepsilon=\pm\Gamma_R/2$. Such a situation corresponds to a single quantum dot attached to a superconducting lead. When $q_s\neq 0$, the four Andreev bound states emerge due to hybridization of the two dots {\it via} the superconducting lead.
However, for sufficiently small values of the parameter $q_s$, only two maxima appear in the DOS as one can notice in Fig.\ref{Fig:3}(a,c). This is because for a small value of $q_s$, the distance between the energies $\varepsilon_{+}$ and $\varepsilon_{-}$ (for the positive or negative energy branches) is small enough compared to the level broadening due to coupling to the ferromagnetic leads. As a consequence, the maxima corresponding to the Andreev states $\varepsilon_{+}$ and $\varepsilon_{-}$ are not resolved and only one peak can be observed at each energy branch. Furthermore, when increasing the parameter $q_s$, these maxima become well resolved and the four peaks emerge in the density of states. With a further increase of the parameter $q_s$, the two peaks corresponding to the energies $\pm\varepsilon_{-}$, approach each other, and for $q_s$ close to 1 they finally merge into a single peak. Similar behavior can be noticed in the transmission displayed in Fig.~\ref{Fig:3}(b,d). However, the situation becomes more complex when $q_s$ approaches 1. Height of the central peak in the transmission diminishes then and the peak finally disappears for $q_s=1$, although the peak at $\varepsilon=0$ in DOS survives even for $q_s=1$. To understand this behavior one should note that the Andreev bound states located at $\pm\varepsilon_{-}$ become degenerate  for $q_s=1$ and $\pm\varepsilon_{-}=0$. The states corresponding to $\pm\varepsilon_{-}$ become then effectively decoupled from the superconducting lead and no transmission is associated with the central peak in the density of states. Such a behavior of the transmission function has a significant influence on the linear conductance, which is determined by transmission at the Fermi level, $\varepsilon =0$.
\begin{figure}
\begin{center}
\includegraphics[width=0.45\textwidth,angle=0]{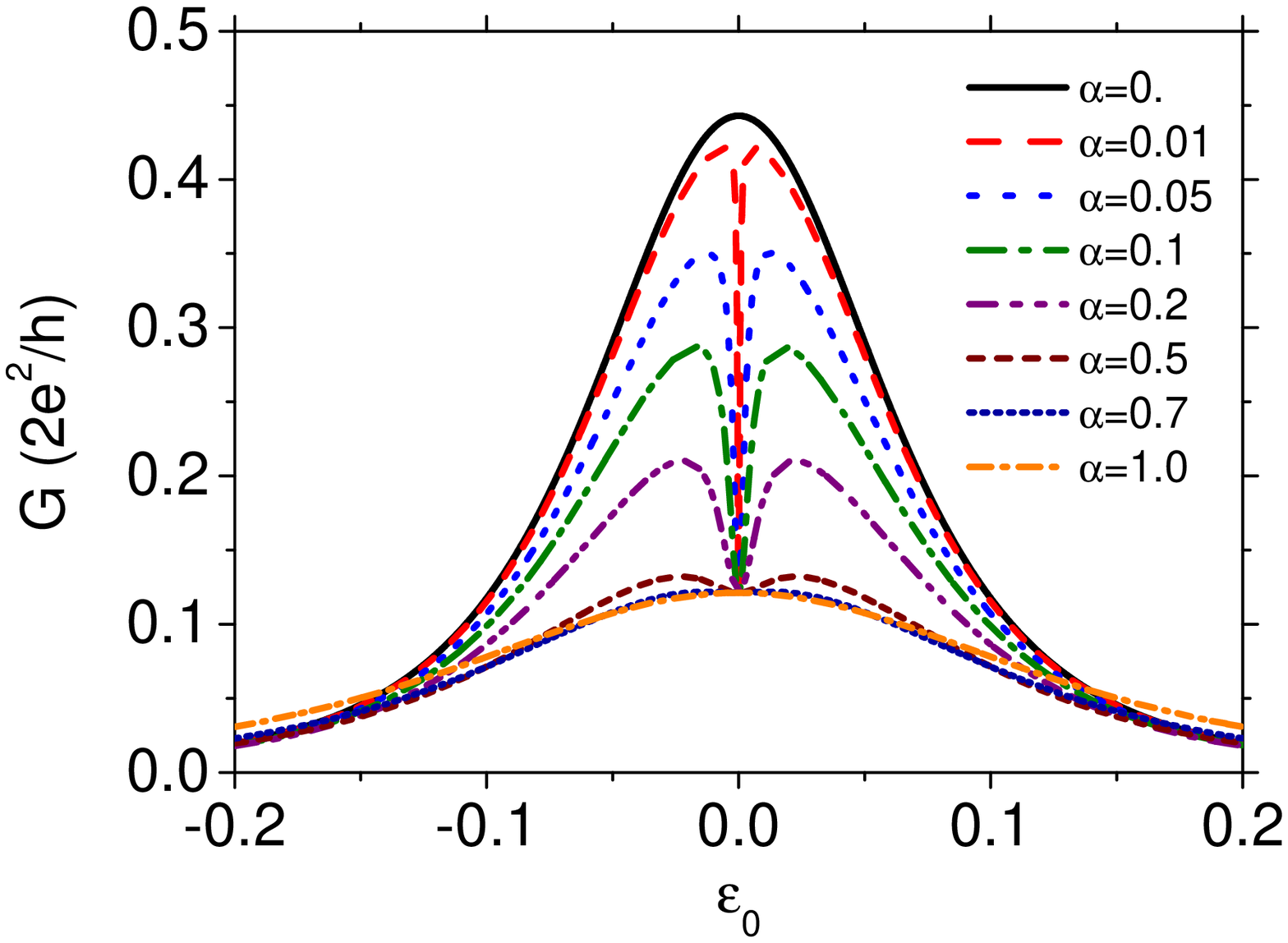}
\caption{\label{Fig:4}
(color on-line) Linear conductance  as a function of the level energy $\varepsilon_0$, calculated for indicated values of the parameter  $\alpha$. The other parameters are: $p=0$, $r=2$,
$\beta=1$, $U=0$.}
\end{center}
\end{figure}

The linear conductance as a function of the dot's level position, $\varepsilon_0$, is shown in Fig.~\ref{Fig:4} for different asymmetries in the coupling of the two dots to the superconductor. This asymmetry is controlled by the  parameter $\alpha$ (when $\alpha=1$, the two dots are coupled to the superconducting lead with equal strengths). The linear conductance in Fig.~\ref{Fig:4} is shown for indicated values of the parameter $\alpha$ and for $q_s=1$. For $\alpha=0$, only one of the two dots is coupled to the superconductor and the conductance as a function of $\varepsilon_0$ reveals a single Lorenzian  peak. When the two dots are equally coupled to the superconducting  lead, $\alpha=1$, the conductance also has a Lorenzian-like shape, but the peak is broader and of lower height. This is due to suppression of the transmission shown in Fig.~\ref{Fig:3}(b).   The situation becomes much more interesting for intermediate values of the parameter $\alpha$, $\alpha\in(0,1)$. A dip structure emerges then in the conductance at $\varepsilon_0=0$, and the local minimum approaches the maximum value of the conductance for $\alpha=1$. As the parameter $\alpha$ decreases, the dip becomes gradually narrower and finally disappears for $\alpha=0$. The dip appears due to a mismatch in the densities of states of the two quantum dots.

\subsubsection{Finite Coulomb repulsion, $U>0$}

\begin{figure}
\begin{center}
\includegraphics[width=0.45\textwidth,angle=0]{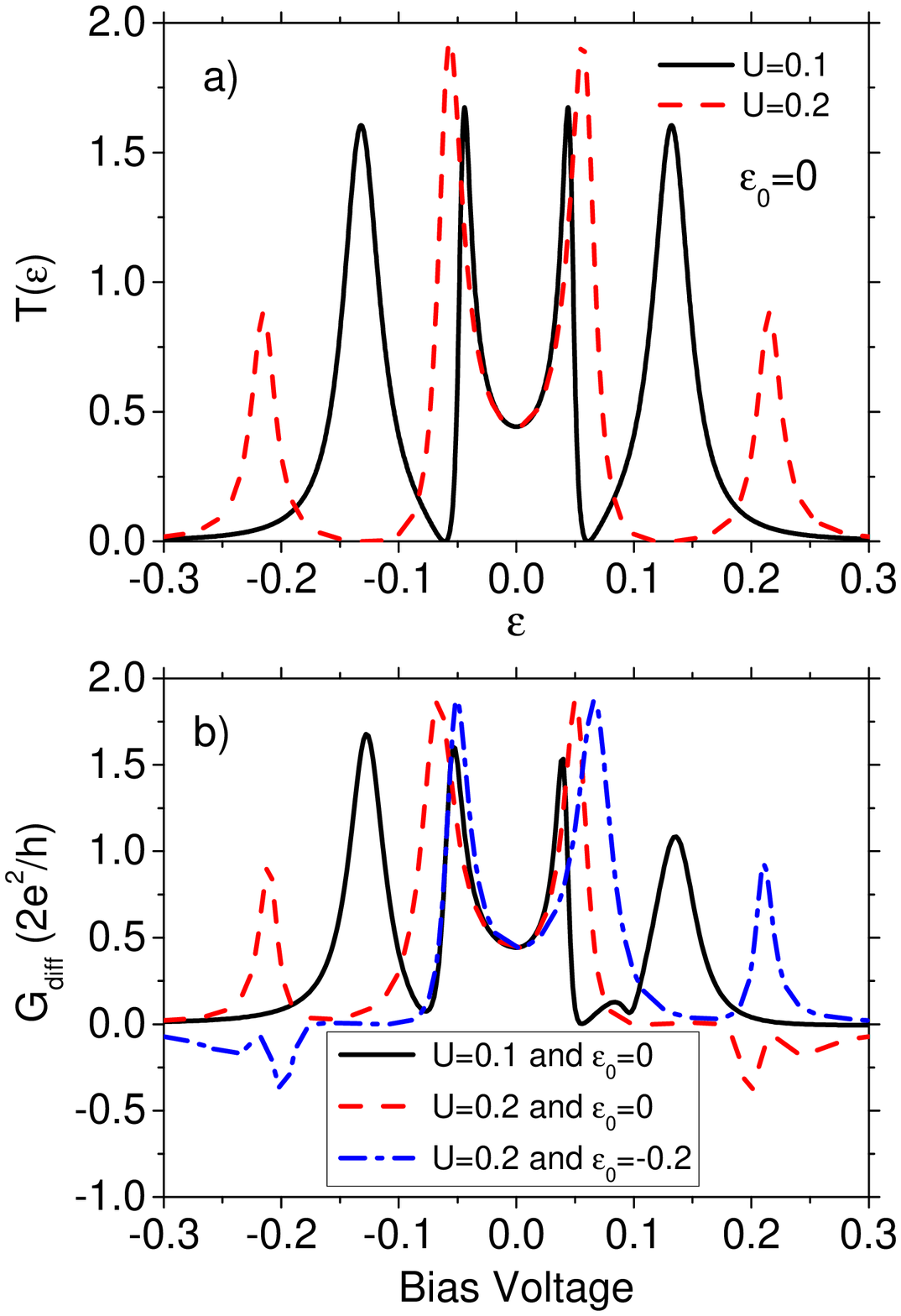}
\caption {\label{Fig:5} (color on-line) Equilibrium transmission function (a), and differential
conductance as a function of applied voltage (b), determined for
indicted values of the intradot Coulomb parameters $U$ and dots' energy levels $\varepsilon_0$. The other
parameters: $p=0$, $r=1$, $q_s=1$, $\alpha=\beta=1$.
}
\end{center}
\end{figure}
\begin{figure}
\begin{center}
\includegraphics[width=0.45\textwidth,angle=0]{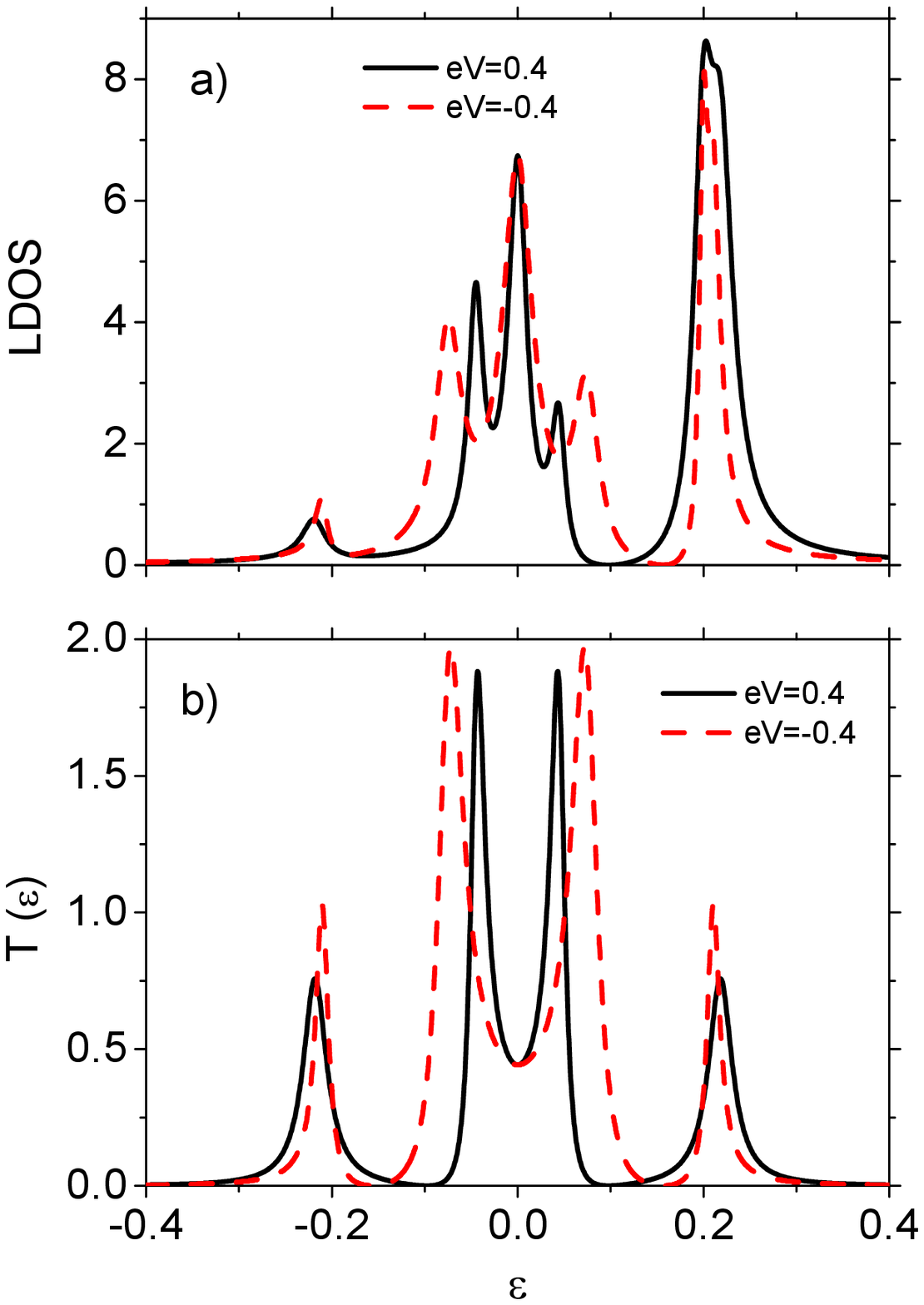}
\caption {\label{Fig:5b} (color on-line) Local density of states (a), and transmission as a function of energy (b), corresponding to  $U=0.2$ and $\varepsilon_0=0$, and calculated for two opposite bias voltages, $eV=0.4$ and $eV=-0.4$. The other
parameters: $p=0$, $r=1$, $q_s=1$, $\alpha=\beta=1$.
}
\end{center}
\end{figure}
In this section we investigate the influence of the intra-dot Coulomb repulsion on the transport characteristics.
In Fig.~\ref{Fig:5}(a) the transmission function at equilibrium is displayed for indicated
values of the Coulomb integral $U$. As one might expect, the Coulomb
interaction leads to doubling of the peak structure. However, the
intensities of the satellite peaks are smaller due to
the lack of electron-hole symmetry for the parameters assumed in Fig.~\ref{Fig:5}a.
This lack of particle-hole symmetry leads to asymmetry in the differential
conductance with respect to the bias reversal, as shown in Fig.~\ref{Fig:5}(b).
For positive voltages, $eV>0$, pairs of electrons coming from normal-metal leads are injected into the superconductor, whereas for $eV<0$ Cooper pairs are extracted from the superconducting lead and the corresponding electrons are then injected into the normal-metal leads.
Interestingly, though the differential conductance reveals the bias reversal asymmetry, the transmission function is still symmetric, see Fig.~\ref{Fig:5}a. However,
the lack of
electron-hole symmetry leads to asymmetric (with respect to
$\varepsilon=0$) density of states of the quantum dots.
This has some consequences for the Andreev current as the particle-hole
symmetry optimizes creation of Cooper pairs, whose energy is equal
to the energy of two electrons and must be equal to the
electrochemical potential of the superconductor, $\mu_R=0$. Accordingly,
the Andreev current is optimized when the electron-hole symmetry exists in the
density of states of the two dots.

An additional feature of the transport characteristics is a negative differential conductance which occurs for higher values of the parameter $U$, see Fig.~\ref{Fig:5}(b). This feature is a consequence of the above mentioned asymmetry in the density of states due to the lack of electron-hole symmetry, and the bias dependence of the transmission function. For $U=0.2$, this asymmetry is
more pronounced than for $U=0.1$, and  leads to negative values of the differential conductance for a finite region of bias voltages. Tuning the gate voltage one may switch the negative differential conductance from positive bias voltage region to negative one, see the curve for $U=0.2$ and $\varepsilon_0=-0.2$  in Fig.~\ref{Fig:5}(b).

To clarify origin of the asymmetry in the differential conductance with respect to the bias reversal, and the appearance of the negative differential conductance, we show in Fig.~\ref{Fig:5b} the transmission and corresponding density of states calculated for the indicated values of bias voltages. In particular, we calculated these quantities for the bias voltage at which the negative differential conductance appears, $eV=0.4$, and for the opposite voltage, $eV=-0.4$. For these values of the bias voltage $eV$ the side peaks in the density of states start to contribute to the Andreev current. Thus, we concentrate only on the energy scales relevant for those peaks. First, we notice that for positive bias voltage, $eV=0.4$, the heights of the transmission peaks (both central and side ones) are smaller than those for negative bias, $eV=-0.4$. The reduction of transmission leads to a reduction of current for positive bias voltage and eventually to the  negative differential conductance. One should also notice that the transmission is symmetric with respect to energy. However, there is no negative differential conductance for negative bias voltages. To get a better  insight into this phenomenon we show in Fig.~\ref{Fig:5b} (a) the corresponding density of states. The densities of states for both dots are equal for the assumed parameters. For the bias voltages considered here, only the side peaks are relevant for the Andreev transport. One can notice that the amplitudes of the satellite peaks, both for $eV=0.4$ and $eV=-0.4$, situated in the positive energy branch are much larger than the amplitudes of the corresponding  peaks on the negative energy side. However, the Andreev reflection processes require transferring electrons through both resonances - one large and one small. Thus, it is the small resonance that determines the  Andreev current. As the amplitude of the small resonance for $eV=-0.4$ is larger than the intensity of the corresponding peak for $eV=0.4$, this explains the suppression of current for  positive voltage.

\begin{figure}
\begin{center}
\includegraphics[width=0.45\textwidth,angle=0]{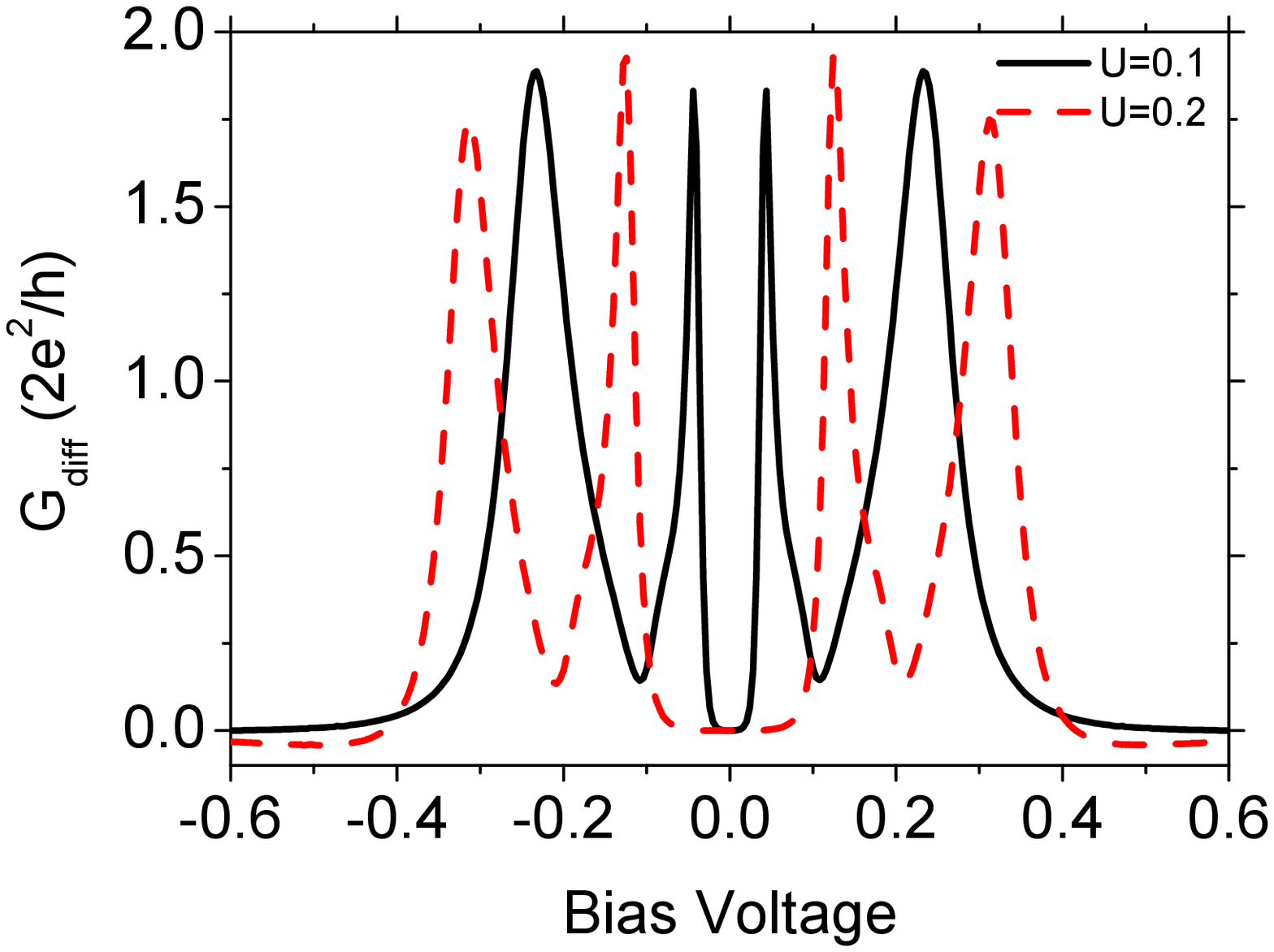}
\caption{\label{Fig:6}
(color on-line) Bias dependence of the differential conductance calculated for indicated values of the Coulomb parameter $U$.
The other parameters are: $p=0$, $r=1$, $\alpha=\beta=1$, and $\varepsilon_0=-U/2$.}
\end{center}
\end{figure}
\begin{figure}
\begin{center}
\includegraphics[width=0.45\textwidth,angle=0]{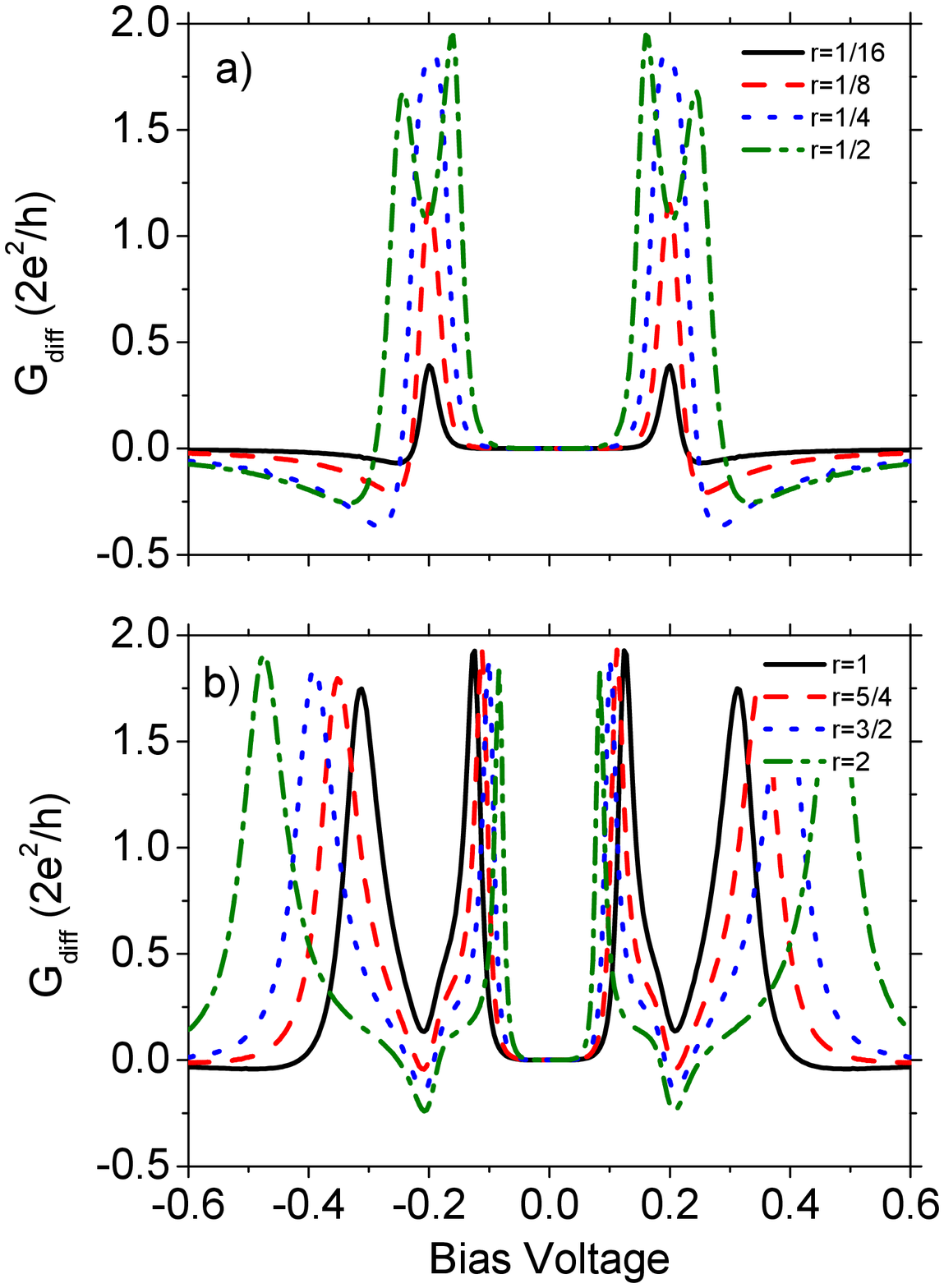}
\caption{\label{Fig:7}
(color on-line) Bias dependence of the differential conductance for indicated values of the asymmetry parameter $r$.
The other parameters are as in Fig.~\ref{Fig:6}.}
\end{center}
\end{figure}

The
electron-hole symmetry exists
if the condition $\varepsilon_0=-U/2$ is satisfied. In Fig.~\ref{Fig:6} we show the differential conductance calculated for indicated values of the Coulomb parameter $U$. Here, due to the above mentioned electron-hole symmetry, the differential conductance is symmetric with respect to the bias reversal. However, regions with negative differential conductance still may emerge. To demonstrate this we display in Fig.~\ref{Fig:7} the differential conductance for different values of the asymmetry parameter $r$.
One can notice that for asymmetry parameter $r\leq 0.25$, the differential conductance consists of two maxima only. Similarly as in the noninteracting case, see Fig.~\ref{Fig:2}, the broadening due to coupling with the ferromagnetic leads is here larger than the separation of the Andreev bound states. With increasing   coupling to the superconducting lead, $r>0.25$, separation between the pair of Andreev levels exceeds the levels' broadening and additional peaks in the differential conductance become resolved.
Apart from this, a negative differential conductance appears. However, it appears  for both positive and negative bias voltages as the differential conductance must be now symmetric with respect to the bias reversal.  Moreover, for $r<1$ the negative differential conductance appears in a broad bias range, whereas for $r>1$ it appears only in a limited region of bias voltages.

\subsection{Ferromagnetic leads, $p>0$}

Consider now  the case when the normal-metal leads are ferromagnetic. Our main objective is to study how ferromagnetism of the leads affects the Andreev tunneling. As before, we start from the noninteracting case, $U=0$.

\subsubsection{Vanishing Coulomb repulsion, $U=0$}

\begin{figure}
\begin{center}
\includegraphics[width=0.47\textwidth,angle=0]{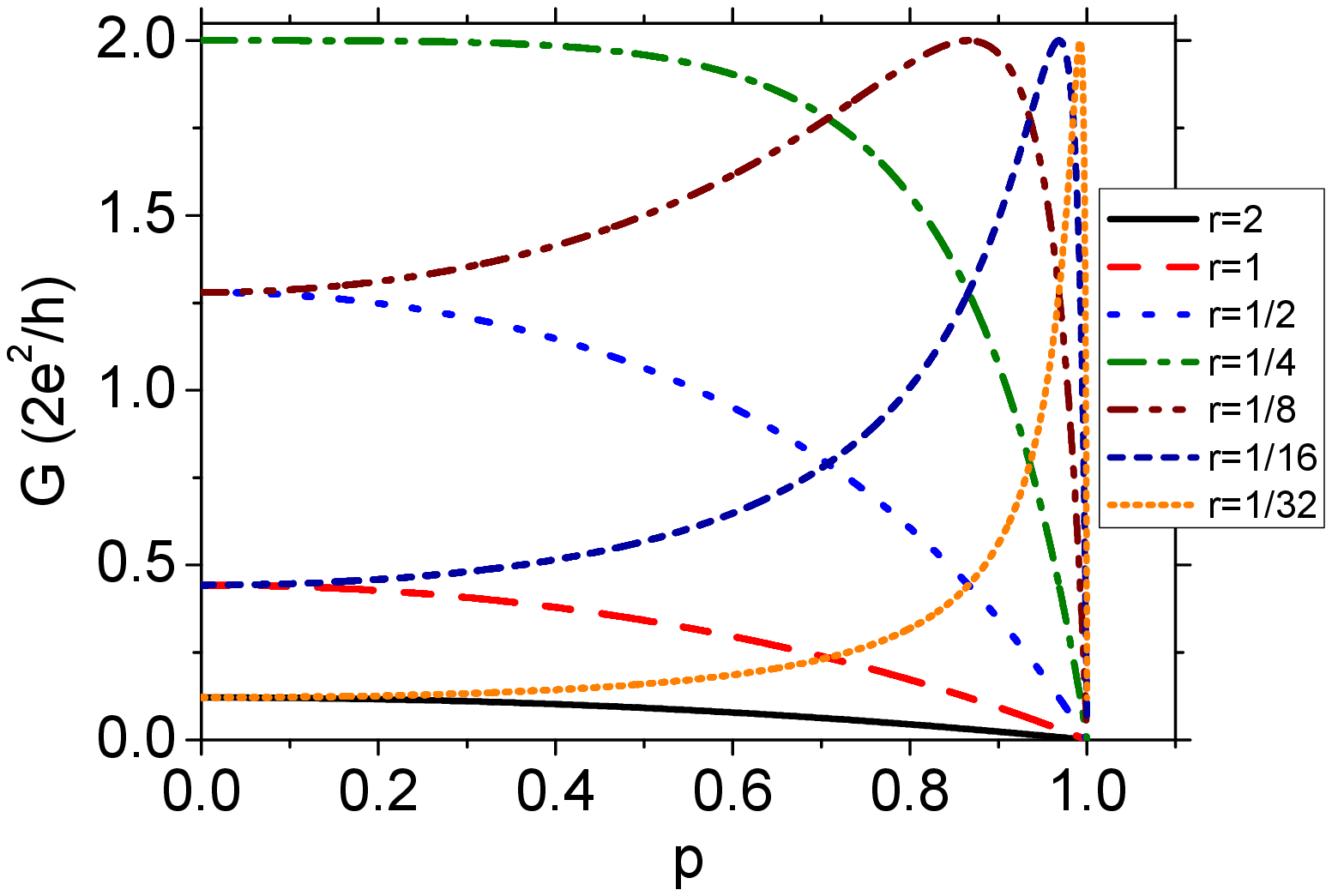}
\caption{\label{Fig:8}
(color on-line) Linear conductance as a function of ferromagnetic leads polarization $p$ calculated for indicated values of the asymmetry parameter $r$.
The other parameters are: $\varepsilon_0=0$, $U=0$, $\alpha=\beta=1$ and $q_s=1$.}
\end{center}
\end{figure}

Figure~\ref{Fig:8} displays
the linear conductance as a function of the polarization factor
$p$ for the parallel magnetic configuration and for different values
of the asymmetry parameter $r$. One could expect that the
conductance should drop  with increasing $p$, as
increasing polarization leads to depletion of electrons with one
 spin orientation. However, Fig.~\ref{Fig:8} clearly shows
that this is true only for $r>1/4$, whereas for $r<1/4$ there is a
certain finite value of the polarization, $p=p_0$, for which the
conductance reaches maximum, $G=4e^2/h$. This maximum is
achieved when the polarization $p$ and the asymmetry coupling
parameter $r$ satisfy the following condition,
$16r^2+p^2=1$. This condition is an analog of a matching condition at the
interface between ferromagnet and superconductor,
$k_{F\uparrow}k_{F\downarrow}=k_{s}^2$, which is expressed in terms
of the corresponding Fermi wave vectors in the ferromagnet ($k_{F\uparrow}$ and $k_{F\downarrow}$) and in the superconductor ($k_{s}$).
For $r>1/4$, this condition
is not satisfied and the conductance drops monotonously with
increasing polarization factor $p$, whereas for $r>1/4$ there
exist such a value of $p$, $p=p_0$, for which the condition is
satisfied and the conductance behaves in a nonmonotonous way
reaching maximum for $p=p_0$. For half-metallic ferromagnets ($p=1$), the
conductance is totally suppressed for any value of the asymmetry
(despite of the trivial case, $r=0$). This is rather obvious as
the Cooper pairs consist of two electrons with opposite spins.

\begin{figure}
\begin{center}
\includegraphics[width=0.47\textwidth,angle=0]{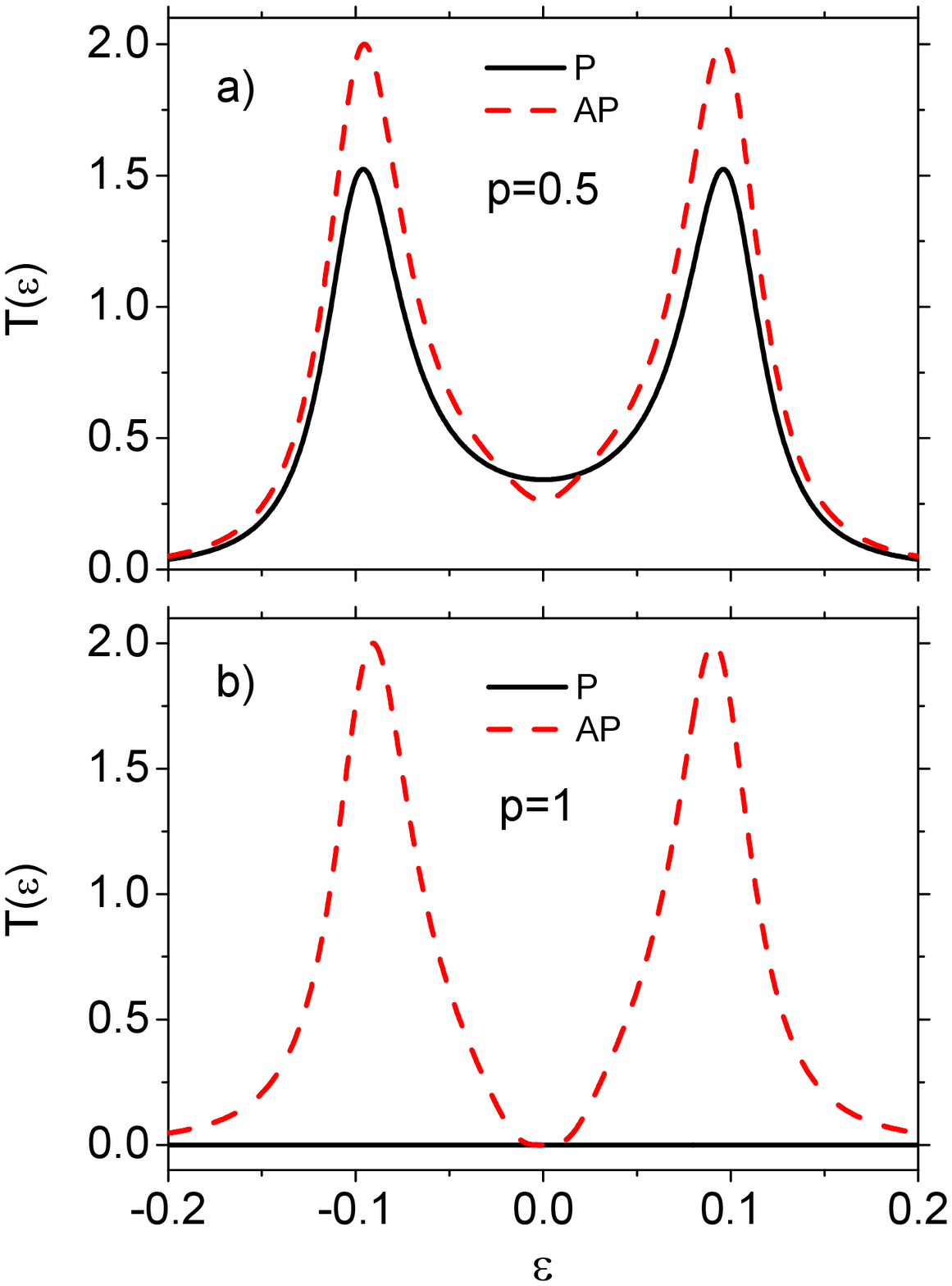}
\caption{\label{Fig:9} (color on-line) Transmission for parallel (P) and
antiparallel (AP) magnetic configurations, and for polarization
(a) $p=0.5$, (b) $p=1$. The other parameters:
$r=1$, $q_s=1$, $\alpha=\beta=1$, $\varepsilon_0=0$,
$U=0$.}
\end{center}
\end{figure}

Change in  magnetic configuration  has a significant impact on
the Andreev transport, too. Figure~\ref{Fig:9} shows that generally
the transmission in the antiparallel magnetic configuration
exceeds that in the parallel one (except for a narrow region of small values of $\varepsilon$). Thus, one
should expect a negative tunnel magnetoresistance for a relatively wide  bias
range. However, for bias voltages close to zero, the
usual behavior can be observed. Moreover, the transmission in the
parallel magnetic configuration does not reach maximum value,
since for $r>1/4$ the matching condition is not satisfied for
any $p$. In the case of half-metallic leads ($p=1$), the transmission
in parallel magnetic configuration is totally suppressed, whereas
transport in the antiparallel configuration can occur for finite
bias voltages. In such a  case, the
 transmission is purely due to the {\it crossed Andreev tunneling}. Thus,
 such a device can be used for experimental verification of the presence and role of CAT processes.

\subsubsection{Finite Coulomb repulsion, $U>0$}

\begin{figure}
\begin{center}
\includegraphics[width=0.35\textwidth,angle=0]{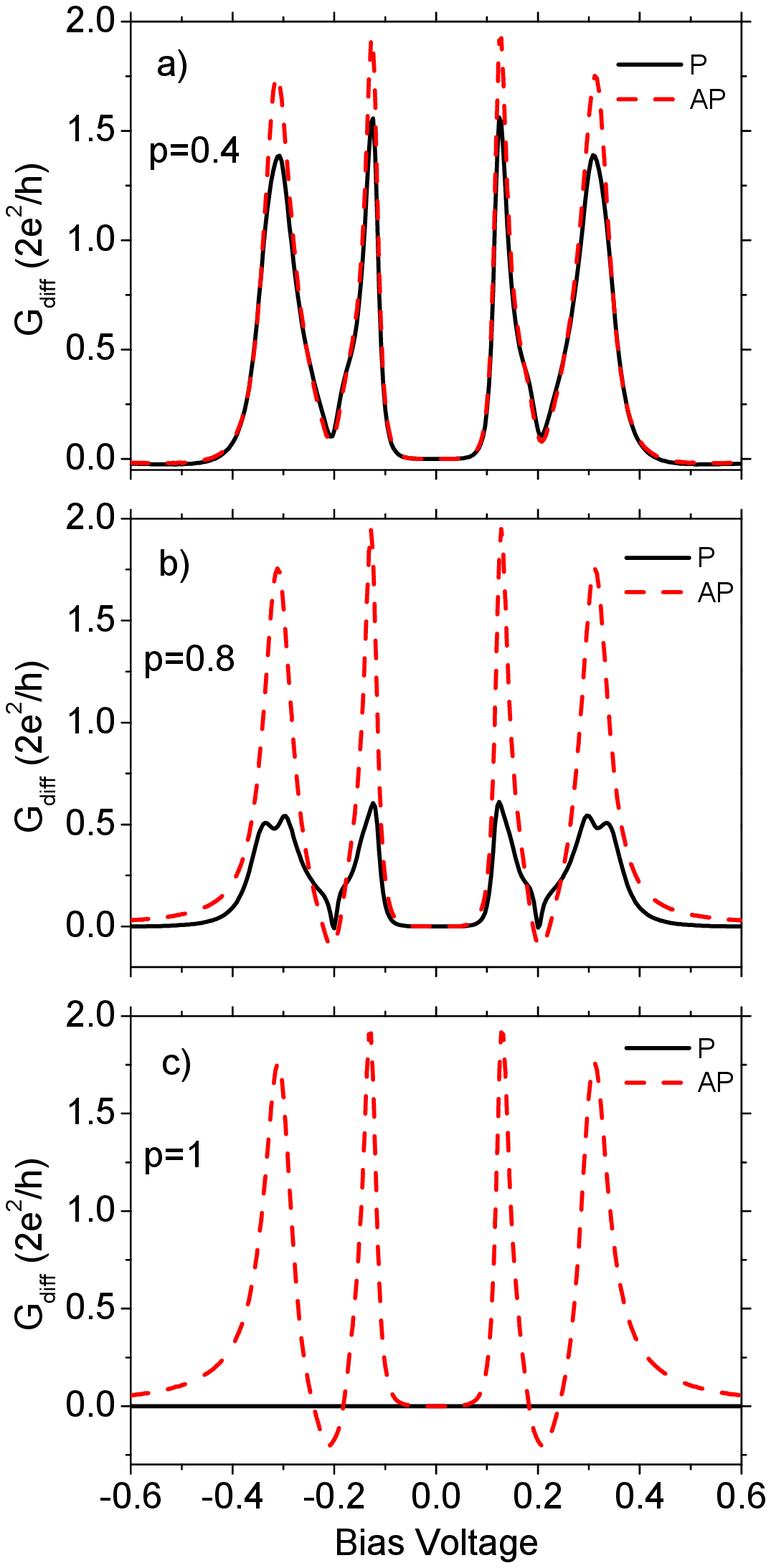}
\caption{\label{Fig:10}
(color on-line) Bias dependence of the differential conductance calculated for indicated values of the spin polarization $p$ and for parallel (P) and antiparallel (AP) magnetic configuration. The other parameters are: $U=-0.2$, $r=1$, $\alpha=\beta=1$, $\varepsilon_0=-U/2$.}
\end{center}
\end{figure}

To complete our study we investigate the influence of Coulomb interaction on the spin-polarized Andreev transport.
Figure~\ref{Fig:10} shows the bias dependence of the differential conductance, calculated for parallel and antiparallel magnetic configurations and for indicated values of the leads' polarization $p$. Similarly as in the noninteracting case ($U=0$), the conductance in parallel magnetic configuration decreases with increasing spin polarization $p$. In turn, the height of the maxima in the differential conductance for the AP configuration are not sensitive to the change in polarization $p$. However, for a sufficiently large polarization $p$, a region of negative differential conductance appears, which becomes more and more pronounced as the factor $p$ tends to its maximal value. This feature is evidently induced by the competition of leads' magnetism and Coulomb interaction as it is absent for $U=0$. However, in the P configuration the negative differential conductance is present only for small values of $p$ and it disappears when $p$ is sufficiently large. Moreover, another feature can be noticed for the P alignment, i.e. splitting of the satellite resonances. In the case of half metallic leads, see Fig.~\ref{Fig:10}(c),  the differential conductance vanishes for parallel  alignment for any bias voltage. For half-metallic leads, the direct Andreev reflection is totally blocked and only CAT processes are responsible for the Andreev current (for the bias voltage assumed here).

\section{Summary}

We have considered Andreev tunneling in a system of two quantum dots coupled to external leads in a fork
geometry. More specifically,  the dots were coupled to a common superconducting lead and each dot was also coupled to
an individual ferromagnetic electrode. By using the Green function method, we calculated the current due to
direct and crossed Andreev tunneling. The key objective was the description and understanding of the effects due to
Coulomb interaction on the dots and leads' ferromagnetism.

The presented results have shown that asymmetry of the coupling
between the dots and electrodes (normal and superconducting) has a
significant influence on transport characteristics.   In
particular, it has been shown that coupling to ferromagnetic leads
gives rise to broadening of the  Andreev reflections, while
magnitude of the coupling to superconducting lead has an influence
on their position. Apart from this, Coulomb interactions on the
dots lead generally to asymmetry in the differential conductance.

A matching condition, taking into account coupling asymmetry and
polarization of ferromagnetic leads, has also been discussed. From
this condition follows that for each value of the polarization
factor $p$  there is an asymmetry parameter $r$, for which the
conductance reaches maximum. From the results follows that even
for large polarization $p$ the conductance may reach $2e^2/h$. Apart
from this, it has been shown that magnetic configuration also has
a significant influence on the transport properties of the DQD
systems.

\appendix

\section{Isolated double dot Green's function}

Dots' Green  function is defined as
$\mathbf{g}^{r}_\sigma(t,t')=-i\theta(t-t')\langle
\{\psi(t),\psi^{\dag}(t')\}\rangle$, with the vector
$\psi\equiv\Psi=(d_{1\sigma}, d_{1\bar{\sigma}}^{\dag},
d_{2\sigma}, d_{2\bar{\sigma}}^{\dag})$. The casual Green's function matrix is block-diagonal and has the following form:
\begin{equation}
\mathbf{g}_\sigma(z)=\left(
                       \begin{array}{cccc}
                         g_{1e}^{\sigma\sigma} & g_{1e}^{\sigma{\bar\sigma}} & 0 & 0 \\
                         g_{1h}^{\bar{\sigma}\sigma} & g_{1h}^{\bar{\sigma}\bar{\sigma}} & 0 & 0 \\
                         0 & 0 & g_{2e}^{\sigma\sigma} & g_{2e}^{\sigma{\bar\sigma}} \\
                         0 & 0 & g_{2h}^{\bar{\sigma}\sigma} & g_{2h}^{\bar{\sigma}\bar{\sigma}} \\
                       \end{array}
                     \right),
\end{equation}
with the
matrix elements $g_{ij}^{\sigma}$:
\begin{equation}
g_{ie/h}^{\sigma\sigma}=\frac{z\mp\varepsilon_{i\sigma}\mp U_i(1-n_{i\bar{\sigma}})}{(z\mp\varepsilon_{i\sigma})(z\mp\varepsilon_{i\sigma}\mp U_i)}
\label{Eq:F1}
\end{equation}
and
\begin{equation}
g_{ie/h}^{\sigma\bar{\sigma}}=\frac{U_i N_{ie/h}}{(z\mp\varepsilon_{i\sigma})(z\mp\varepsilon_{i\sigma}\mp U_i)},
\label{Eq:F2}
\end{equation}
where the upper sign refers to electron functions, $\langle\langle
d_{i\sigma}|d_{j\sigma'}^{\dag}\rangle\rangle$ and $\langle\langle
d_{i\sigma}|d_{j\sigma'}\rangle\rangle$, whereas the lower sign
corresponds to  hole functions, $\langle\langle d_{i\sigma}^{\dag}|d_{j\sigma'}\rangle\rangle$ and
$\langle\langle d_{i\sigma}^{\dag}|d_{j\sigma'}^{\dag}\rangle\rangle$.

For the
electron functions $N_{ie}=\langle d_{i\bar{\sigma}}d_{i\sigma}\rangle$, whereas for the hole functions, $N_{ih}=\langle d_{i\bar{\sigma}}^{\dag}d_{i\sigma}^{\dag}\rangle$
To obtain retarded Green function one sets $z=\varepsilon+i0^{+}$.
In numerical calculations we dropped off-diagonal terms of $\mathbf{g}_\sigma(\epsilon)$ putting $N_{ie}=N_{ih}\approx 0$. Although, these correlations (straightforwardly connected with proximity effect) are nonzero for small values of Coulomb repulsion, they rather weakly influence the transport characteristics in the considered regime.
\begin{acknowledgements}
This work was supported from the
`Iuventus Plus' project No. IP2011 059471 for years 2012-2014.
P.T. also acknowledges support from the Ministry of Science and Higher
Education as a research project N N202 169536 in years 2009-2011 and by the European Union under European Social Fund Operational Programme "Human Capital" (POKL.04.01.01-00-133/09-00 ).
\end{acknowledgements}

\end{document}